# Secondary electron emissions and dust charging currents in the nonequilibrium dusty plasma with power-law distributions

Jingyu Gong and Jiulin Du [a]

*Department of Physics, School of Science, Tianjin University, Tianjin 300072, China*

**Abstract**

We study the secondary electron emissions induced by the impact of electrons on dust grains and the resulting dust charging processes in the nonequilibrium dusty plasma with power-law distributions. We derive new expressions of the secondary emitted electron flux and the dust charging currents that are generalized by the power-law $q$-distributions, where the nonlinear core functions are numerically studied for the nonextensive parameter $q$. Our numerical analyses show that the power-law $q$-distribution of the primary electrons has a significant effect on the secondary emitted electron flux as well as the dust charging currents, and this effect depends strongly on the ratio of the electrostatic potential energy of the primary electrons at the dust grain's surface to the thermodynamic energy, implying that a competition in the dusty plasma between these two energies plays a crucial role in this novel effect.

[a] E-mail: jiulindu@yahoo.com.cn

## I. INTRODUCTION

Dusty plasma is ubiquitous in the space and terrestrial environments. It is composed of normal electron-ion plasma and additional charged dust grains immerged in the plasma. Compared with the electron and the ion, the mass and size of dust grain are much greater and therefore increase the complexity of dusty plasma.[1,2] The dust charging process is one of the complex problems in the dusty plasma physics, which strongly depends on the interactions between dust grains, as well as dust grains and the other particles such as electrons, ions, and the background radiation. The elementary processes that lead to the dust charging include the collection of plasma particles, the secondary electron emission, the photoemission, the thermionic emission, and so on.[3] In dusty plasma, the secondary electrons can be released from the surface of a dust grain when the energetic primary particles (incident particles), e.g. electrons or ions, are hitting on the surface.[4-6] The secondary electron emission could be significant in dusty plasma if the energy of these primary particles becomes sufficiently large. For the secondary electron emission induced by the impact of the electrons on the dust grains, the secondary yield is defined as the ratio of the emitted electron flux and the primary electron flux. A semi-rational formula of the secondary yield is usually given as a function of the energy $E$ of the primary electron,[4,5]

$$\delta(E) = 7.4 \, \delta_M \frac{E}{E_M} \exp\left(-2\sqrt{E/E_M}\right), \tag{1}$$

where $\delta_M$ is the maximum value of $\delta$ and there is $E = E_M$ when the secondary yield gets its maximum. The quantities $\delta_M$ and $E_M$ are determined by materials of dust grains, e.g. $\delta_M \approx 1$ for metals and semiconductors, and $\delta_M \approx 2\sim30$ for insulators. In general, $E_M$ is about 300~2000eV. Usually, the dust charging processes were studied on the basis of traditional statistical theory, the results of which have been widely applied to explain some of the relevant physical phenomena observed in dusty plasmas.[7,8]

The dust charging is a nonequilibrium process taking place in the dusty plasma systems far away from equilibrium, such as the solar winds, the planetary magnetospheres, magnetosheaths, etc. In many physical situations of both space and terrestrial nonequilibrium palsmas, e.g., the laboratory plasma with a steep temperature gradient,[9] the earth's plasma sheets and the solar winds containing plentiful superthermal particles,[10] the solar interior,[11] the planetary magnetospheres and magnetosheaths,[12] it has been often observed that the statistical properties of particles may not be described by the Maxwellian-Boltzmann distribution, but can be described by the power-law distributions. The typical forms of such power-law distributions (see Ref.13 and the references therein), which have attracted great attention, include the $\kappa$-distributions or the generalized Lorentzian distributions observed in the solar winds and space plasmas, the $q$-distributions investigated for complex systems within nonextensive statistical mechanics, and those $\alpha$-distributions noted in physics, chemistry and elsewhere like $P(E) \sim E^{-\alpha}$. Theoretically, by introducing new fluctuation-dissipation relations, one has accurately produced many different forms of the power-law distributions from the stochastic dynamics of Langevin equations.[13] In recent years, nonextensive statistics has been employed to investigate some of basic physical characteristics of the plasmas with power-law distributions, such as ion acoustic waves[14-16], dust acoustic waves[17,18], electron acoustic waves[19], dust charging processes[20,21], and Jeans' instability in space plasma[22] etc. In these new attempts, the nonextensive effect has been considered and interpreted as the nature of a nonequilibrium stationary state of the systems with Coulombian long-range interactions.[23] It is worth mentioning that what distinguishes an equilibrium state from a nonequilibrium stationary state is not the form of a stationary distribution, but the presence or absence of the detailed balance. The Maxwellian velocity $q$-distribution has been known to represent a nonequilibrium stationary state of the long-range interacting systems.[24] Nevertheless, according to the present understanding,[25,26] the usually employed $q$-distribution function is not factorized for kinetic and potential energies because there is no consideration of the nonextensivity of the energy, and it is found to be just an isothermal distribution and therefore it might not be a correct description of a nonequilibrium stationary state.

The reader might be also interested in some of the recent astrophysical applications of the power-law $q$-distributions,[26, 27] where the nonextensivity (physical meaning of the nonextensive parameter $q \neq 1$) has been presented a real physics regarding the non-isothermal nature and non-uniform velocity dispersion in the astrophysical systems with self-gravitating long-range interactions.

In the previous work,[20] we studied the dust charging processes in the collections of electrons and ions in the nonequilibrium dusty plasma with power-law distributions. In this work, we will study the secondary electron emissions induced by the impact of primary electrons and the resulting dust charging currents in the nonequilibrium dusty plasma with power-law distributions. The paper is organized as follows. In Sec.II, we study the secondary emitted electron flux due to the impacts of the primary electrons on the surface of dust grains when the electrons obey the power-law $q$-distribution. In Sec.III, we study the dust charging currents including the secondary electron current in the dusty plasma. In all these works, the numerical analyses are made for the relevant physical quantities with different nonexetensive parameters of the electrons. Finally, in Sec.IV, we give the conclusions and discussions.

## II. THE SECONDARY EMITTED ELECTRON FLUX AND POWER-LAW DISTRIBUTION

We consider the spatially homogeneous dusty plasma with power-law velocity distributions. Thus the particle number densities of ions and electrons are constants. The dust grain of radius $r_d$ may be considered as a spherical probe immersed in the plasma. The secondary electron emissions are induced by the impact of the primary electrons on the dust grains. Usually, as the radius of the dust grain is much smaller than the plasma Debye radius, and it is also far less than the average distance between dust grains, the orbit-limited motion (OLM) method can be used to calculate the electron flux on the surface of dust grains. In this way, the cross section for collisions between the dust grains and the electrons is given[1] by

$$\sigma_d = \pi r_d^2 \left(1 + \frac{e\phi_d}{E_k}\right), \tag{2}$$

where the surface potential of a dust grain is defined, relative to the plasma potential, by $\phi_d = Q_d/r_d$ with the charge $Q_d$. And the primary electron flux per unit area on the surface of dust grain can be generally expressed[5, 6] by

$$J_e = \frac{1}{4\pi r_d^2} \int_0^{2\pi} d\varphi \int_0^{\pi} \sin\theta d\theta \int_{v_{\min}}^{v_{\max}} \sigma_d f(v_e) v_e^3 dv_e$$

$$= \int_{\varepsilon_{\min}}^{\varepsilon_{\max}} \left(1 + \frac{e\phi_d}{\varepsilon}\right) \frac{2\pi\varepsilon}{m_e^2} f(\varepsilon) d\varepsilon, \tag{3}$$

where $m_e$ is the mass of electron, $\varepsilon$ is the kinetic energy of the electron, and $f(\varepsilon)$ is the distribution function of the electrons. We consider the nonequilibrium dusty plasma with the power-law distributions that can be described by nonextensive statistics. In that way, the distribution function of the electrons[26] is

$$f(\varepsilon) = n_e A_q \left(\frac{m_e}{2\pi k_B T_e}\right)^{3/2} \left[1 - (1-q_e)\frac{\varepsilon}{k_B T_e}\right]^{1/(1-q_e)}, \tag{4}$$

where $k_B$ is the Boltzmann constant, $n_e$ is the number density, $T_e$ is the temperature, $q_e$ is the nonextensive parameter, and $A_q$ is the normalization coefficient,

$$A_q = \begin{cases} \dfrac{1}{4}(5-3q_e)(3-q_e)\sqrt{1-q_e}\, \Gamma\left(\dfrac{1}{1-q_e}+\dfrac{1}{2}\right) \Big/ \Gamma\left(\dfrac{1}{1-q_e}\right), & \text{for } q_e < 1, \\ \dfrac{1}{2}(5-3q_e)\sqrt{q_e-1}\, \Gamma\left(\dfrac{1}{q_e-1}\right) \Big/ \Gamma\left(\dfrac{1}{q_e-1}-\dfrac{1}{2}\right), & \text{for } q_e > 1. \end{cases} \tag{5}$$

When one sets $q_e \to 1$, Eq.(4) becomes the Maxwell-Boltzmann distribution. The physical meaning of the nonextensive parameter is explained as follows. In the nonequilibrium plasma, $q_e \neq 1$ can be associated with the relevant physical quantities by the equation[23]

$$k_B \nabla T_e = (1-q_e)e\nabla\varphi, \tag{6}$$

where the potential function $\varphi$ is determined by the Poisson's equation. It is clear that the nonextensive parameter is $q_e \neq 1$ if and only if the temperature gradient is $\nabla T_e \neq 0$. Thus Eq.(4) is a stationary-state distribution in the nonequilibrium plasma.[23]

Substituting Eq.(4) into Eq.(3), the primary electron flux becomes

$$J_e = n_e A_q \left(\frac{m_e}{2\pi k_B T_e}\right)^{3/2} \frac{2\pi}{m_e^2} \int_0^{\varepsilon_{\max}} (\varepsilon + e\phi_d) \left[1 - (1-q_e)\frac{\varepsilon}{k_B T_e}\right]^{\frac{1}{1-q_e}} d\varepsilon. \tag{7}$$

Now we consider the secondary emitted electron flux $J_s$. The expression of $J_s$ depends on the nature of the surface potential of dust grains. If the surface potential is $\phi_d < 0$, then the secondary emitted electron flux[5, 6] is $J_s = \int \delta(E) dJ_e$. Using Eqs.(1) and (7), one has

$$J_s = n_e A_q \left(\frac{m_e}{2\pi k_B T_e}\right)^{\frac{3}{2}} \frac{2\pi}{m_e^2} \int_0^{\varepsilon_{\max}} \delta(\varepsilon + e\phi_d) \left[1 - (1-q_e)\frac{\varepsilon}{k_B T_e}\right]^{\frac{1}{1-q_e}} (\varepsilon + e\phi_d) d\varepsilon. \tag{8}$$

The upper limit of this integral is $\varepsilon_{\max} = k_B T_e / (1-q_e)$ if $q_e < 1$, and $\varepsilon_{\max} = \infty$ if $q_e > 1$. Further, Eq.(8) can be expressed by

$$J_s = 3.7 \delta_M \sqrt{\frac{k_B T_e}{2\pi m_e}} n_e F_-^q (U, x), \tag{9}$$

where the core function $F_-^q(U, x)$ is

$$F_-^q(U, x) = A_q x^2 \int_0^{u_{\max}} u^5 \left[1 - (1-q_e)(xu^2 - U)\right]^{\frac{1}{1-q_e}} e^{-u} du, \tag{10}$$

with $U = e\phi_d / k_B T_e$ and $x = E_M / 4k_B T_e$, and one has $u_{\max} = \infty$ if $q_e > 1$ and $u_{\max} = \sqrt{[1+(q_e-1)U]/x(1-q_e)}$ with $U \geq -1/(1-q_e)$ if $q_e < 1$. The dimensionless parameters $\delta_M$ and $x$ are typically[5] $\delta_M < 30$ and $x \simeq 25 \sim 500$.

On the other hand, if the surface potential is $\phi_d > 0$, not all the secondary electrons are able to escape from the surfaces of dust grains, and therefore the secondary emitted electron flux[6] is $J_s = (1 + e\phi_d / k_B T_s) e^{-e\phi_d / k_B T_s} \int \delta(E) dJ_e$, where $T_s$ is the temperature of the secondary electrons. Namely,

$$J_s = n_e A_q \left(\frac{m_e}{2\pi k_B T_e}\right)^{3/2} \frac{2\pi}{m_e^2}\left(1+\frac{e\phi_d}{k_B T_s}\right)e^{-e\phi_d/k_B T_s}$$

$$\times \int_0^{\varepsilon_{\max}} \delta(\varepsilon + e\phi_d)\left[1-(1-q_e)\frac{\varepsilon}{k_B T_e}\right]^{\frac{1}{1-q_e}}(\varepsilon + e\phi_d)d\varepsilon. \qquad (11)$$

The upper limit of this integral is $\varepsilon_{\max} = k_B T_e/(1-q_e)$ if $q_e < 1$, and $\varepsilon_{\max} = \infty$ if $q_e > 1$. Further, in the same way as Eq.(9), Eq.(11) can be expressed by

$$J_s = 3.7\delta_M \sqrt{\frac{k_B T_e}{2\pi m_e}} n_e \left(1+\frac{U}{\sigma_s}\right)e^{-U/\sigma_s} F_+^q(U,x) \qquad (12)$$

with $\sigma_s = T_s/T_e$, where the core function $F_+^q(U,x)$ is

$$F_+^q(U,x) = A_q x^2 \int_{\sqrt{U/x}}^{u_{\max}} u^5 \left[1-(1-q_e)(xu^2-U)\right]^{\frac{1}{1-q_e}} e^{-u} du. \qquad (13)$$

The upper limit of the integral is $u_{\max} = \sqrt{[(q_e-1)U+1]/x(1-q_e)}$ if $q_e < 1$, and is $u_{\max} = \infty$ if $q_e > 1$.

It is clear that the secondary emitted electron flux, Eq.(9) or Eq.(12), depends strongly on the nonlinear core function, $F_-^q(U,x)$ or $F_+^q(U,x)$, and hence on the nonextensive parameter $q_e \neq 1$. If we take the limit $q_e \to 1$, these functions will recover the forms with the Maxwellian-Boltzmann distribution,[5] i.e. $F_-^1(U,x) = e^U F_5(x)$ and $F_+^1(U,x) = e^U F_{5,B}(x)$. In order to show dependence of the new nonlinear core functions (therefore the secondary emitted electron flux) on the nonextensive parameter more clearly, $F_-^q(U,x)$ and $F_+^q(U,x)$ have been studied numerically on the basis of Eqs.(10) and (13) and they have been illustrated in Fig.1, respectively, for several different values of $q_e$. Fig.1 shows that the power-law distribution ($q_e \neq 1$) of primary electrons has a significant effect on both core functions. When $x$ is small, the effect on $F_-^q(-1, x)$ is very significant, but on $F_+^q(1, x)$ is very small. With the increase of $x$, the effect on $F_-^q(-1, x)$ decreases slowly, but on $F_+^q (1, x)$ increases gradually and becomes significant. And then, when $x$ becomes large enough, the effect on both $F_-^q (-1, x)$ and $F_+^q (1, x)$ will be smaller and smaller gradually.

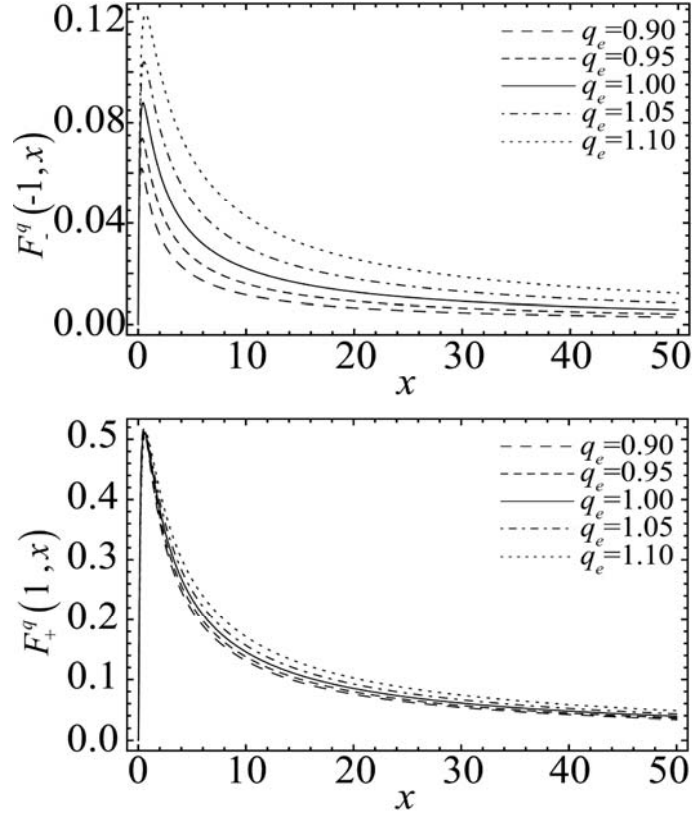

FIG.1. The core functions $F_-^q (-1, x)$ and $F_+^q (1, x)$ for different $q_e$.

In Fig.2, the secondary emitted electron flux $J_s$, as a function of $U$, has been illustrated for several different $q_e$ by using a normalized secondary emitted electron flux defined by $\tilde{J}_s = J_s / \sqrt{n_e^2 k_B T_e / 2\pi m_e}$. These numerical analyses were carried out based on Eq.(9) for the part of $U < 0$, and Eq.(12) for the part of $U > 0$, in which $x$ was fixed at $x=45.6$ and other two physical parameters were typically taken as $\delta_M = 15$ and $\sigma_s = 1.5$. Fig.2 shows that the power-law distribution of the primary electrons ($q_e \neq 1$) has a significant effect on the secondary emitted electron flux $J_s$ only if the value of $U$ is about -6 ~ 4, but it has almost no any effect on $J_s$ if the value of $U$ is beyond 4. And if the value of $U$ is less than -6, with the decrease of $U$ this effect also become gradually smaller and smaller. We conclude that the effect of the power-law distribution ($q_e \neq 1$) on the secondary emitted electron flux depends strongly on the value of $U = e\phi_d/k_B T_e$, a ratio of the electron's potential energy at the surface of dust grains to their thermodynamic energy. Only if $|U|$ is not large, can the above effect be significant, implying that a competition between these two energies of the primary electrons in the nonequilibrium dusty plasma plays an important role in this effect.

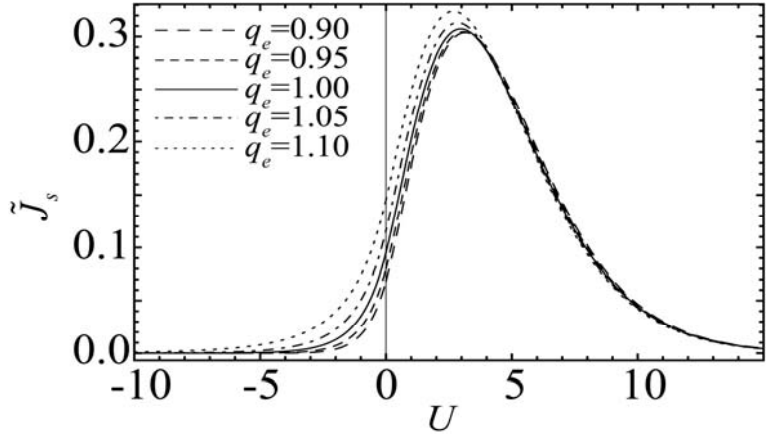

FIG.2. The secondary emitted electron flux as a function of $U$ for different $q_e$. Other parameters are fixed at $x = 45.6$, $\delta_M = 15$ and $\sigma_s = 1.5$.

## III. THE DUST CHARGING CURRENTS AND POWER-LAW DISTRIBUTION

In our physical consideration, the dust charging currents should include the secondary electron charging current, $I_s$, the electron charging current, $I_e$, and the ion charging current, $I_i$. The secondary current $I_s$ can be calculated by the relation $I_s = 4\pi r_d^2 e \cdot J_s$, where the secondary emitted electron flux $J_s$ has been given by Eqs.(9) and (12). Thus, the total dust charging current is $I_T = I_i + I_e + I_s$. In order to simplify the calculations, we may introduce a normalized dust charging current,

$$\tilde{I}_T = I_T \Big/ 4\pi r_d^2 ne\sqrt{k_B T_e/2\pi m_e} \, . \tag{14}$$

Then the current balance equation is $\tilde{I}_T = 0$. In plasma, the quasi-neutral condition, $n_e/n_i = 1 - Z_d n_d/n_i$, should be satisfied. For the case of isolated dust grains, $Z_d n_d/n_i$ is much smaller than 1, and so, in the following calculations, we assume the number density to be $n_i \approx n_e \equiv n$.

If the surface potential of dust grains is $\phi_d < 0$, the electron charging current and the ion charging current are, respectively,[20]

$$I_e = -4\pi r_d^2 enB_{qe}\sqrt{\frac{k_B T_e}{2\pi m_e}}\left[1+(1-q_e)\frac{e\phi_d}{k_B T_e}\right]^{(3-2q_e)/(1-q_e)}, \tag{15}$$

and

$$I_i = 4\pi r_d^2 enB_{qi}\sqrt{\frac{k_B T_i}{2\pi m_i}}\left[1-(3-2q_i)\frac{e\phi_d}{k_B T_i}\right], \tag{16}$$

where $q_i$ is the nonextensive parameter of ions, $T_i$ is the temperature of ions and $m_i$ is the mass. The normalization coefficient $B_{qj}$ ($j = i$ for ions and $j=e$ for electrons) is given by

$$B_{qj} = \begin{cases} \dfrac{(5-3q_j)(3-q_j)\sqrt{1-q_j}}{4(2-q_j)(3-2q_j)}\Gamma\left(\dfrac{1}{1-q_j}+\dfrac{1}{2}\right)\Big/\Gamma\left(\dfrac{1}{1-q_j}\right), & \text{for } q_j < 1, \\[2mm] \dfrac{(5-3q_j)\sqrt{q_j-1}}{2(2-q_j)(3-2q_j)}\Gamma\left(\dfrac{1}{q_j-1}\right)\Big/\Gamma\left(\dfrac{1}{q_j-1}-\dfrac{1}{2}\right), & \text{for } q_j > 1. \end{cases} \tag{17}$$

The second electron charging current can be written by using Eq.(9) as

$$I_s = 14.8\pi r_d^2 e\,\delta_M n\sqrt{\frac{k_B T_e}{2\pi m_e}}F_-^q(U,x) \, . \tag{18}$$

Using Eqs.(15), (16) and (18), one finds

$$\tilde{I}_T = B_{qi}\sqrt{\frac{\sigma_i}{\mu}}\left[1-(3-2q_i)\frac{U}{\sigma_i}\right] - B_{qe}\left[1+(1-q_e)U\right]^{\frac{3-2q_e}{1-q_e}} + 3.7\delta_M F_-^q(U,x), \tag{19}$$

where we have set $U = e\phi_d/k_B T_e$, $x = E_M/4k_B T_e$, $\mu = m_i/m_e$ and $\sigma_i = T_i/T_e$. If we take the nonextensive parameters $q_e \to 1$ and $q_i \to 1$, these functions will recover the forms in the Maxwell-Boltzmann distributions.

If the surface potential of dust grains is $\phi_d > 0$, the electron charging current and the ion charging current are, respectively,[20]

$$I_e = -4\pi r_d^2 e n B_{qe} \sqrt{\frac{k_B T_e}{2\pi m_e}} \left[1 + (3-2q_e)\frac{e\phi_d}{k_B T_e}\right], \tag{20}$$

and

$$I_i = 4\pi r_d^2 e n B_{qi} \sqrt{\frac{k_B T_i}{2\pi m_i}} \left[1 - (1-q_i)\frac{e\phi_d}{k_B T_i}\right]^{(3-2q_i)/(1-q_i)}. \tag{21}$$

While the second electron charging current can be written by using Eq.(12) as

$$I_s = 14.8\pi r_d^2 e \, \delta_M n \sqrt{\frac{k_B T_e}{2\pi m_e}} \left(1 + \frac{e\phi_d}{k_B T_s}\right) e^{-e\phi_d/k_B T_s} F_+^q(U,x). \tag{22}$$

Using Eqs.(20)-(22), one finds

$$\tilde{I}_T = B_{qi}\sqrt{\frac{\sigma_i}{\mu}} \left[1 - (1-q_i)\frac{U}{\sigma_i}\right]^{\frac{3-2q_i}{1-q_i}} - B_{qe}\left[1 + (3-2q_e)U\right]$$

$$+ 3.7\delta_M \left(1 + \frac{U}{\sigma_s}\right) e^{-U/\sigma_s} F_+^q(U,x), \tag{23}$$

where $\sigma_s = T_s/T_e$. If we take the nonextensive parameters $q_e \to 1$ and $q_i \to 1$, these functions also recover the forms in the Maxwell-Boltzmann distributions.

Since we are here concerned with the dust charging current generated by the secondary electron emissions induced mainly by the impact of the electrons on the dust grains, the power-law distribution of the ions might be unimportant. In the following numerical analyses, in order to simplify the calculations, we might set the nonextensive parameter of ions $q_i \to 1$. In this case, Eq.(19) becomes

$$\tilde{I}_T = \sqrt{\frac{\sigma_i}{\mu}}\left(1 - \frac{U}{\sigma_i}\right) - B_{qe}\left[1 + (1-q_e)U\right]^{\frac{3-2q_e}{1-q_e}} + 3.7\delta_M F_-^q(U,x), \tag{24}$$

and Eq.(23) reduces to

$$\tilde{I}_T = \sqrt{\frac{\sigma_i}{\mu}}e^{-U/\sigma_i} - B_{qe}\left[1 + (3-2q_e)U\right] + 3.7\delta_M\left(1 + \frac{U}{\sigma_s}\right)e^{-U/\sigma_s} F_+^q(U,x). \tag{25}$$

In Fig.3. (a)~(c), the total dust charging current $I_T$, as a function of $U$, has been illustrated for several different values of $q_e$ by using the normalized charging current $\tilde{I}_T$. These numerical analyses were carried out based on Eq.(24) for the part of $U<0$ and Eq.(25) for the part of $U>0$. In Fig.3.(a), the physical parameters were typically taken as $x$=45.6, $\sigma_i=1$ and $\sigma_s=1.5, \delta_M=15$. As we expected, the figure shows that the electrons' power-law distribution ($q_e \neq 1$) has a significant effect on the dust charging current. And it also has a very significant effect on the dust grains' equilibrium potentials, since the equilibrium potentials are determined by $\tilde{I}_T$=0, i.e. the intersections of the curves and the line of $\tilde{I}_T$=0 in the figure. Again, only if $|U|$ is not large, can these effects be more significant.

Nevertheless, the total dust charging current depends not only on the electrons' power-law distribution ($q_e \neq 1$), but also on several dimensionless parameters such as $x$, $\delta_M$, $\sigma_i$ and $\sigma_s$. In Fig.3. (b), we illustrated the same physical situations as that in Fig.3.(a) when $x$=45.6 is replaced by $x$=57, and in Fig.3.(c) when $\delta_M=15$ is replaced by $\delta_M=1.5$. We find that the effects observed in the figures (b) and (c) have become quite different from those in the figure (a). All these analyses showed the vicissitude and complexity of dust charging processes in the nonequilibrium dusty plasma with power-law distributions.

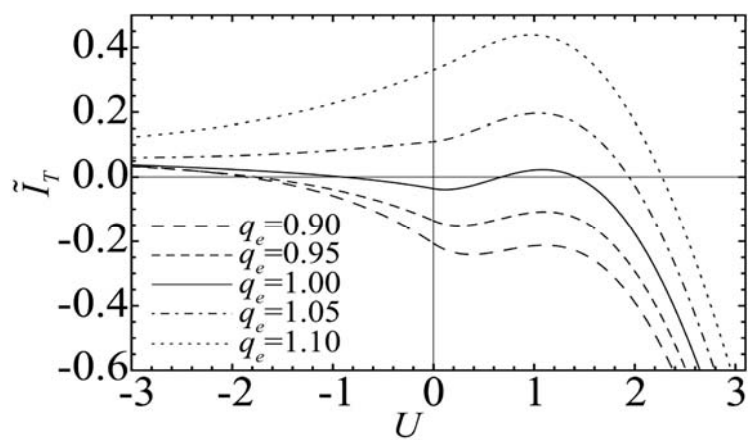

(a). $\tilde{I}_T(U, q_e, x=45.6, \sigma_i=1, \sigma_s=1.5,$ and $\delta_M=15)$

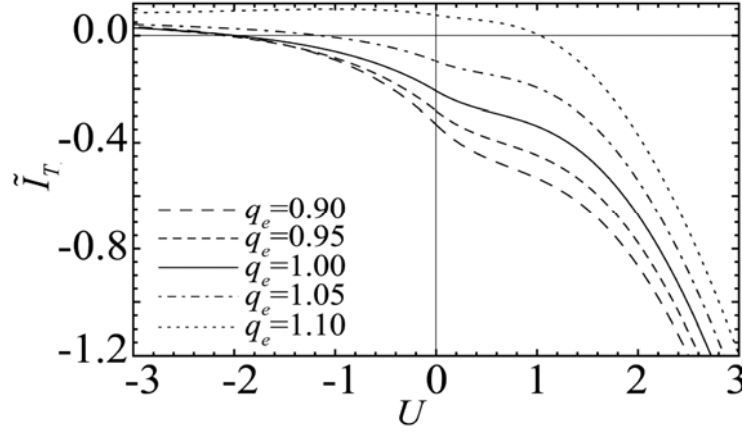

(b). $\tilde{I}_T(U, q_e, x=57, \sigma_i=1, \sigma_s=1.5, \text{ and } \delta_M=15)$

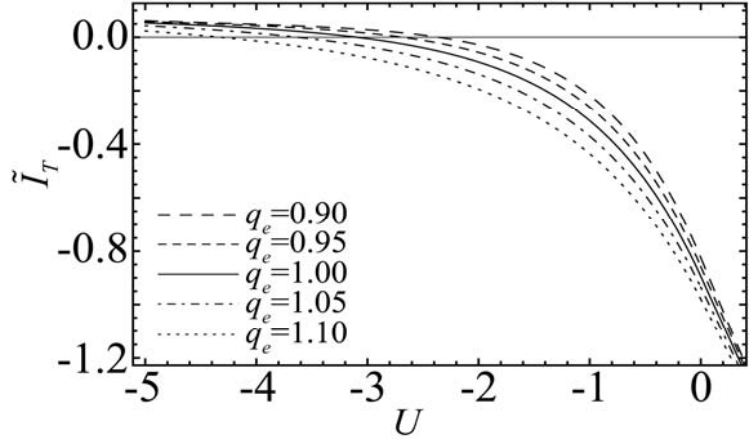

(c). $\tilde{I}_T(U, q_e, x=45.6, \sigma_i=1, \sigma_s=1.5, \text{ and } \delta_M=1.5)$

FIG.3. (a)~(c): The dust charging current $\tilde{I}_T$ is given as a function of $U$ for different $q_e$, where other parameters are fixed. These figures show the effects of $q_e \neq 1$ on $\tilde{I}_T$ and the equilibrium potentials.

## IV. CONCLUSIONS AND DISCUSSIONS

In conclusion, we have studied the secondary electron emissions induced by the impact of the primary electrons on the surface of dust grains and the resulting dust charging processes in the nonequilibrium dusty plasma with power-law distributions. The primary electrons in the dusty plasma were considered to obey the power-law $q$-distribution that can be described by nonextensive statistics. In this case, we derived new expressions of secondary emitted electron flux, i.e. Eq.(9) for $\phi_d < 0$ and Eq.(12)

for $\phi_d > 0$, which depend on the nonlinear core functions $F_-^q(U,x)$ and $F_+^q(U,x)$, respectively. The previous expressions based on the Maxwell-Boltzmann distribution were generalized by the power-law distribution. Our numerical analyses showed that the power-law distribution ($q_e \neq 1$) of the primary electrons had a significant effect on the secondary emitted electron flux, and this effect depended strongly on the ratio of the electrostatic potential energy of the electrons at the surface of dust grains to the thermodynamic energy, implying that a competition in the dusty plasma between these two energies of the electrons plays a crucial role in this effect.

Using the new expressions of secondary emitted electron flux with the power-law $q$-distribution, we have discussed the resulting dust charging currents in the nonequilibrium dusty plasma. The total dust charging current was presented in the form of the power-law distributions. As we expected, the numerical analyses showed that the power-law distribution ($q_e \neq 1$) of the electrons had a very significant effect on the dust charging current and the dust grains' equilibrium potentials. Nevertheless, because this effect also depends on other several dimensionless parameters, these numerical analyses enable us to gain an insight into the vicissitude and complexity of dust charging processes in the nonequilibrium dusty plasma with power-law distributions.

As an example of the discussions, our model of dust charging currents might be applied to calculate the equilibrium potential of rings of Saturn. E ring is the outmost one of the rings of Saturn, which has extremely wide range extending from 3R$_s$ to 8~9.5R$_s$ (R$_s$=6.033× 10$^4$ km is the radius of Saturn) and mostly consists of water ice grains, much like the dusts.[28] According to the observations of Voyager 1, the plasma parameters, temperature and number density, at the radial distance L=6R$_s$ in the E ring[29] are $T_e = 5.0\,\text{eV}$, $n_e = 27\,\text{cm}^{-3}$, $T_{O^+} = 100\,\text{eV}$, $n_{O^+} = 25\,\text{cm}^{-3}$, $T_{H^+} = 16\,\text{eV}$, and $n_{H^+} = 2\,\text{cm}^{-3}$, and there is the best fit of the experimental secondary yield with $\delta_M = 2.35$ and $E_M = 340\,\text{eV}$. Based on these parameters, we can determine the $q_e$-dependent equilibrium potential of the dusty plasma in the E ring with power-law distributions. In fact, many physical characteristics in nonequilibrium complex dusty plasmas are associated with the dust charging processes including the secondary electron emission etc.[30-32] In particular, when the statistical nature of particles in the complex plasmas is not a Maxwellian distribution but the power-law distributions, the new characteristics of the plasmas can be studied by the methods developed in our works.

**ACKNOWLEDGMENTS**

This work is supported by the National Natural Science Foundation of China under Grant No.10675088 and No.11175128, and the Higher School Specialized Research Fund for Doctoral Program under Grant No. 20110032110058.